\documentclass[12pt,preprint]{aastex}

\shorttitle{HD 189733b's Radius and Density}
\shortauthors{Baines \& van Belle et al.}

\begin{document}

\title{Direct Measurement of the Radius and Density of the Transiting Exoplanet HD~189733B with the CHARA Array}

\author{Ellyn K. Baines}
\affil{Center for High Angular Resolution Astronomy, Georgia State University, P.O. Box 3969, Atlanta, GA 30302-3969}
\email{baines@chara.gsu.edu}

\author{Gerard T. van Belle}
\affil{Michelson Science Center, California Institute of Technology, 770 S. Wilson Avenue, MS 100-22, Pasadena, CA 91125}
\email{gerard@ipac.caltech.edu} 

\author{Theo A. ten Brummelaar, Harold A. McAlister}
\affil{Center for High Angular Resolution Astronomy, Georgia State University, P.O. Box 3969, Atlanta, GA 30302-3969}
\email{theo@chara-array.org, hal@chara.gsu.edu}

\author{Mark Swain}
\affil{Jet Propulsion Laboratory, 4800 Oak Grove Drive, Pasadena, CA 91109} 
\email{mark.swain@jpl.nasa.gov}

\author{Nils H. Turner, Laszlo Sturmann, Judit Sturmann}
\affil{Center for High Angular Resolution Astronomy, Georgia State University, P.O. Box 3969, Atlanta, GA 30302-3969}
\email{nils, sturmann, judit@chara-array.org}

\altaffiltext{1}{For preprints, please email baines@chara.gsu.edu.}

\begin{abstract}
We have measured the angular diameter of the transiting extrasolar planet host star HD~189733 using the CHARA O/IR interferometric array. Combining our new angular diameter of $0.377 \pm 0.024$~mas with the Hipparcos parallax leads to a linear radius for the host star of $0.779 \pm 0.052$~$R_{\rm Sun}$ and a radius for the planet of $1.19 \pm 0.08$~$R_{\rm Jup}$. Adopting the mass of the planet as derived by its discoverers, we derive a mean density of the planet of $0.91 \pm 0.18$~g~cm$^{-3}$. This is the first determination of the diameter of an extrasolar planet through purely direct means.
\end{abstract}


\section{Introduction}

A handful of extrasolar planets transit their host stars, causing a reduction in stellar flux as the planet blocks part of the star's disk.  The planet orbiting HD~189733 is one of the fourteen known transiting planets.  Using radial velocity and photometric measurements made at the Haute-Provence Observatory, \citet {2005A&A...444L..15B} discovered a hot Jupiter-like planet with an orbital period of 2.219 days and estimated the star's radius to be 0.76~$\pm$~0.01~$R_\odot$. This value, along with a planet-to-star radius ratio of 0.172~$\pm$~0.003, led to a planetary radius of 1.26~$\pm$~0.03~$R_{\rm Jup}$. More recently, \citet{2006ApJ...650.1160B} refined the orbital parameters using $BVRI$ multi-band photometry and found the planet's radius to be 1.154~$\pm$~0.032~$R_{\rm Jup}$.

We observed HD 189733 using Georgia State University's Center for High Angular Resolution Astronomy (CHARA) Array in order to directly determine the host star's radius and thereby calculate, in a strictly geometric manner, the radius and density for the planet.

Planetary densities were previously estimated from photometric observations of the transiting planets and range from 0.38~g~cm$^{-3}$ for HD~209458b \citep{2000ApJ...529L..45C} to 1.17~g~cm$^{-3}$ for HD~149026b \citep{2005ApJ...633..465S}. These density calculations are highly dependent on estimated stellar diameters based on spectral energy distribution (SED) fits using published photometric values, which are fundamentally indirect in nature, relying upon \emph{a priori} assumptions regarding the host stars' stellar atmospheres. For the four `bright' $(V<12)$ transit host stars, these angular sizes are in the range of 0.05 to 0.40 milliarcseconds (mas). The longest baselines of the CHARA Array are capable of resolving the largest and brightest of these objects.

\section{Interferometric Observations and Diameter Determination}

\subsection{Observations and Data Reduction}
Spatially resolved observations of HD 189733 were obtained with the CHARA Array, a six-element interferometer located on Mount Wilson, California \citep{2005ApJ...628..439M}. The Array operates in two wavelength regimes: in visible wavelengths (470-800 nm) for tracking and tip/tilt correction; and in the near infrared $K'$ (2.13~$\mu$m) and $H$ (1.67~$\mu$m) bands for fringe detection. Because of the small angular diameter for the star, only the $H$-band observations obtained at our longest baseline pair (telescopes E1 and S1) are used in our final diameter analysis.

\begin{figure}[!t]
  \centering \includegraphics[width=1.0\textwidth]
  {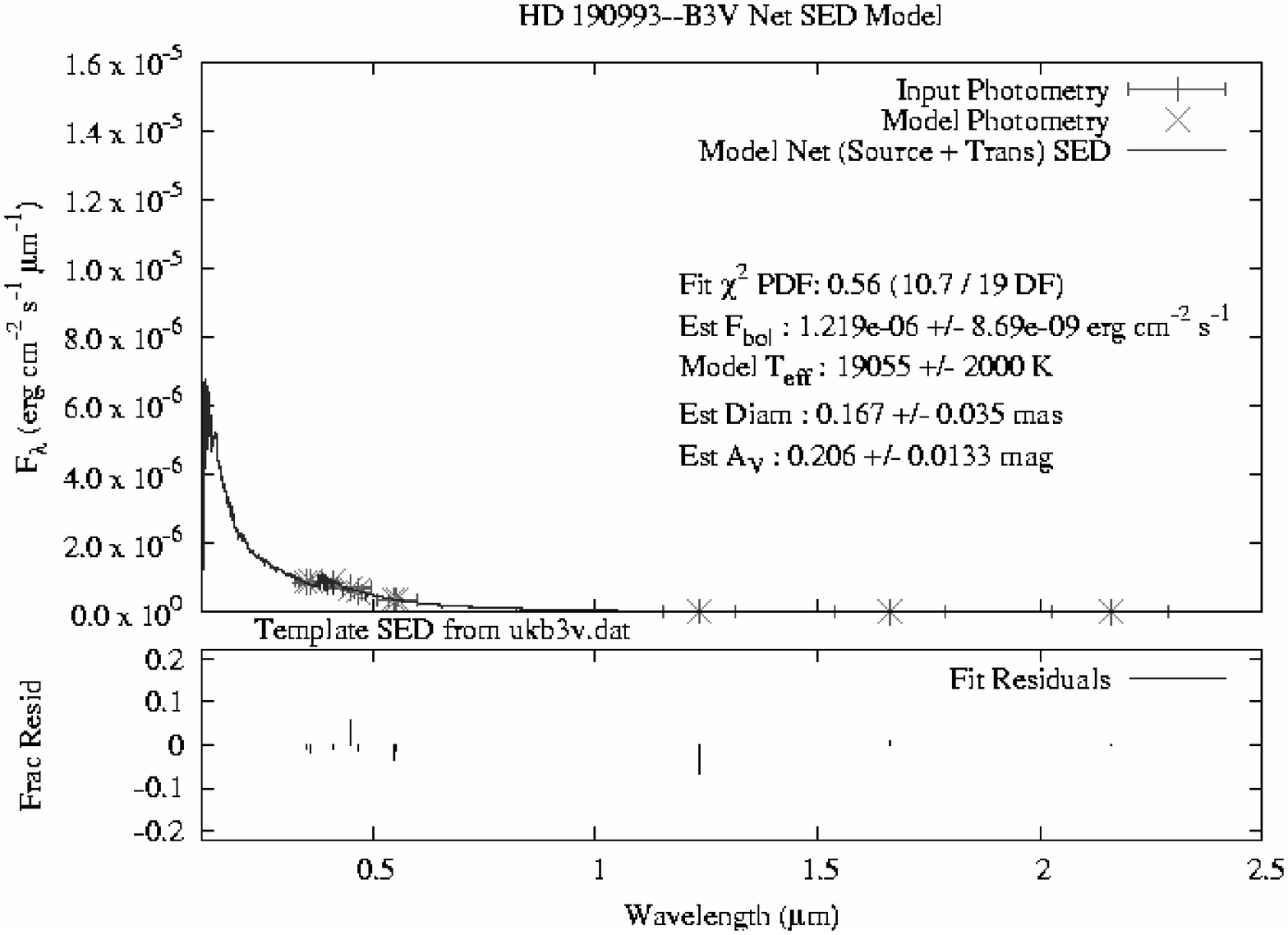}\\
  \caption{SED fit of our calibrator star, HD 190993. In the top panel, vertical bars represent the errors for the data points they overlay and the horizonal bars represent the bandpass of the data point. In the bottom panel, the fractional residuals (difference between data point and fit point, normalized by that data point) are shown for each data point.}
  \label{HD190993_SED}
\end{figure}

HD~189733 was observed on several nights during the summer of 2006 along with the calibrator star HD~190993, a B3~V star offset by 1.7$^{\circ}$, selected on the basis of its small estimated angular diameter and its apparent lack of any close companion. The latter criterion was verified by a thorough literature search while a spectral energy distribution fit to HD~190993 led to an estimated angular diameter of $0.167 \pm 0.035$ mas with no residuals suggestive of a companion (see Figure \ref{HD190993_SED}). This results in a predicted visibility ($V$) for the calibrator of $V_{\rm cal}$=0.961$^{+0.019}_{-0.008}$ at our longest baseline of 330~m, resulting in a contribution of $\sigma_V \simeq 0.01-0.02$ to the calibrated visibility errors seen in Table 1. The small angular size and high visibility of the calibrator means HD~190993 is essentially unresolved using the CHARA Array, and the uncertainty in visibility due to calibrator diameter error is small compared to the measurement error. Therefore uncertainties in the calibrator diameter will not affect the HD~189733 diameter measurement significantly \citep{2005PASP..117.1263V}. Even HD~190993's considerable $v$~sin~$i$ does not contribute error to our diameter fits due to its small angular size.

We note that the M-dwarf companion to HD~189733 reported by \citet{2006ApJ...641L..57B} on the basis of common space motion at an angular separation of 11.2 arcsec is well outside the interferometric field of view, and its presence has no effect on our results. Although the effect on visibility would be small in the first lobe of the $V(B)$ curve, we have confirmed that our observed epochs do not occur within the predicted times of planetary transit or eclipse using the period and reference time of central transit of \citet{2006ApJ...650.1160B}

All our observations were obtained with the single-baseline, pupil-plane ``CHARA Classic'' beam combiner, and we employed the standard practice of observing the target and calibrator sequentially to provide a series of time-bracketed observations from which the instrumental visibilities could be reduced to calibrated values for the target star. The observing practice and reduction process employed here is identical to that described by \citet{2005ApJ...628..453T}. The results of this process are summarized in Table \ref{visibility}.

Single baseline Michelson stellar interferometers measure complex visibilities, usually recorded as amplitudes and phases, which are related to the intensity distribution of the target through a Fourier transform; phase information is typically corrupted by the atmosphere, leaving the amplitude, referred to simply as the $visibility$ \citep{1999ApJ...510..505C}.

\subsection{Diameter Fit}
\begin{figure}[!t]
  \centering \includegraphics[width=1.0\textwidth]
  {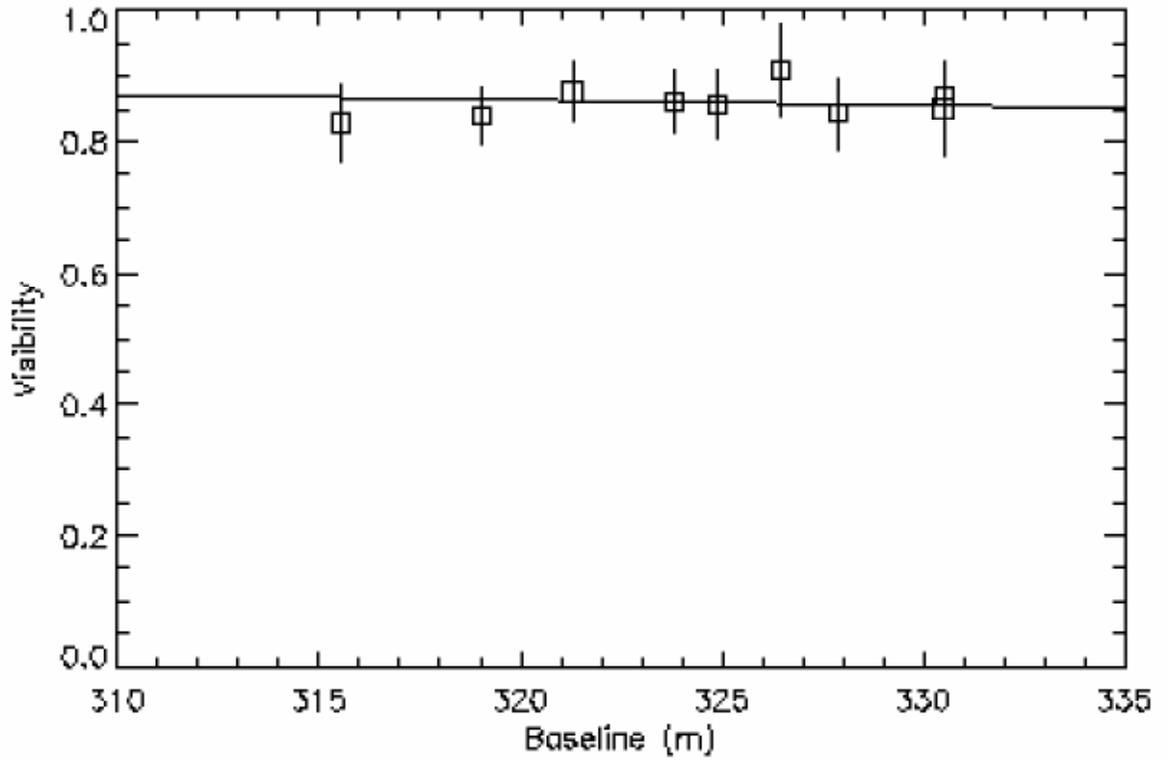}\\
  \caption{The $\chi^2$ fit to HD 189733's visibilities. The solid line represents a theoretical visibility curve for a star with a limb-darkened diameter of 0.376~mas, the boxes are the measured visibilities, and the vertical lines are the measured errors.}
  \label{HD189733_viscurve}
\end{figure}

Diameter fits to the visibilities and baselines from Table \ref{visibility} were performed using the uniform disk (UD) approximation given by:
\begin{equation}
V =  \frac{2 J_1(\rm x) }{\rm x},
\end{equation}
where \emph{J}$_1$ is the first-order Bessel function and
\begin{equation}
\rm x = \pi \emph{B} \theta_{UD} \lambda^{-1},
\end{equation}
where \emph{B} is the projected baseline at the star's position, $\theta_{\rm UD}$ is the apparent UD angular diameter of the star, and $\lambda$ is the wavelength of the observation. The limb-darkened (LD) relationship incorporating the 
linear limb darkening coefficient $\mu_{\lambda}$ \citep{1974MNRAS.167..475B} is given by:
\begin{equation}
V = \left( {1-\mu_\lambda \over 2} + {\mu_\lambda \over 3} \right)^{-1}
\times
\left[
(1-\mu_\lambda) {J_1(\rm x) \over \rm x} + \mu_\lambda {\left( \frac{\pi}{2} \right)^{1/2} \frac{J_{3/2}(\rm x)}{\rm x^{3/2}}} 
\right].
\end{equation}
These fits resulted in $\Theta_{\rm UD}$~=~$0.366 \pm 0.024$ mas and $\Theta_{\rm LD}$~=~$0.377 \pm 0.024$ mas, the latter incorporating $\mu_{\lambda}$~=~0.36 taken from \citet{1995A&AS..114..247C} after adopting log~$g$~=~4.5 and $T_{\rm eff}$~=~5000~K for HD~189733 (see Figure \ref{HD189733_viscurve}). The reduced $\chi^2$ minimization in both cases yielded a value of 1.593, and the errors quoted are for an increase of the $\chi^2$ value of 1.0, that is, the 68\% confidence interval. Dividing this $\chi^2$ by the number of degrees of freedom, which in our case is 8, yields 0.199, which is much less than 1.0 showing that the fit is quite good and that 
our error estimates for the visibility points are conservative. If we rescale these errors bars to force $\chi^2$ to be equal to the number of degrees of freedom, which assumes that there are no systematics in the measurements, they are approximately half the size as they are shown in 
Figure 2 and would also reduce our final error estimates by a factor of two. However, we will remain conservative and continue to use the error estimate based on the raw $\chi^2$ value.

\subsection{Estimate of the Angular Size of HD~189733}

An {\it a priori} estimate of the angular size of HD~189733 is a parameter of considerable interest, because the size of HD~189733b is determined only relative to the size of its parent star from the photometric transit timing data. \citet{2006ApJ...650.1160B} consider no less than four separate methods in their investigation of the system: $V-K$ color angular radius prediction \citep{2004A&A...426..297K}, temperature radius, isochrone radii from \citet{2002A&A...391..195G} and \citet{1998A&A...337..403B}, and the Johnson V - 2MASS $T_{\rm EFF}$ calibration of \citet{2006A&A...450..735M}.

None of these approaches appears to have much merit, since the only primary data we have been able to find in the literature were Tycho $B_T$, $V_T$ \citep{2000PASP..112..961B}, Str{\"o}mgren {\it ubvy} \citep{1993A&AS..102...89O}, and 2MASS $JHK$ photometry \citep{2003tmc..book.....C}. No spectroscopy or measures of log $g$ appear to be available in the literature, nor do direct measures of Johnson photometry. The values of $V$ used in \citet{2004A&A...426..297K} and \citet{2006A&A...450..735M} appear to have been extrapolated from $V_T$. Furthermore, sustained long-term observations of HD~189733 by the {\it MOST} asteroseismology satellite have found the star to be photometrically variable \citep{rowe2007}, casting significant doubt on any radius derived from a photometric relationship. Based on this information, we consider the size errors for HD~189733 quoted in \citet{2005A&A...444L..15B} and \citet{2006ApJ...641L..57B} derived from the methods cited above to be underestimates.

\begin{figure}[!t]
  \centering \includegraphics[width=0.5\textwidth]
  {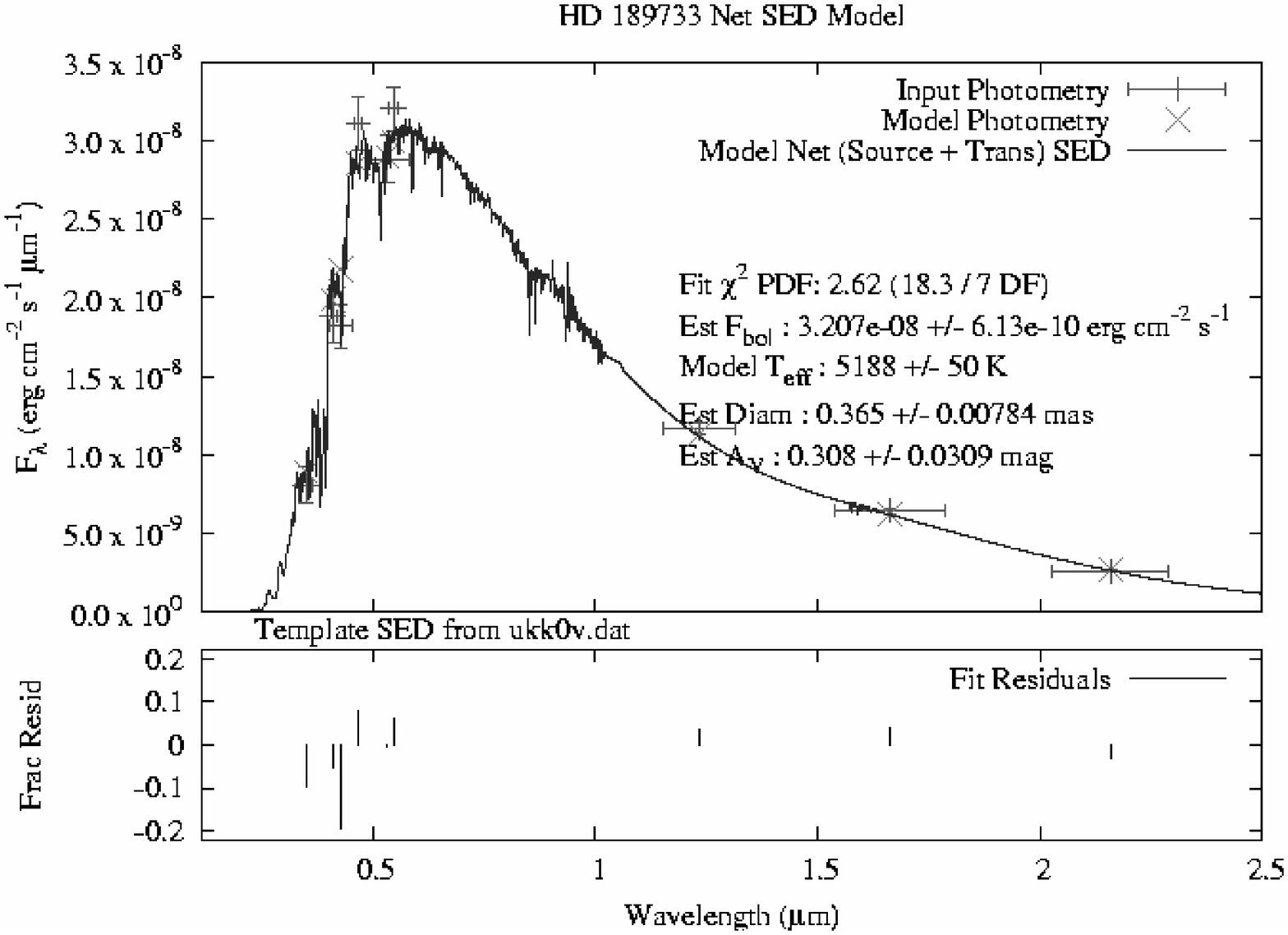}\\
  \centering \includegraphics[width=0.5\textwidth]
  {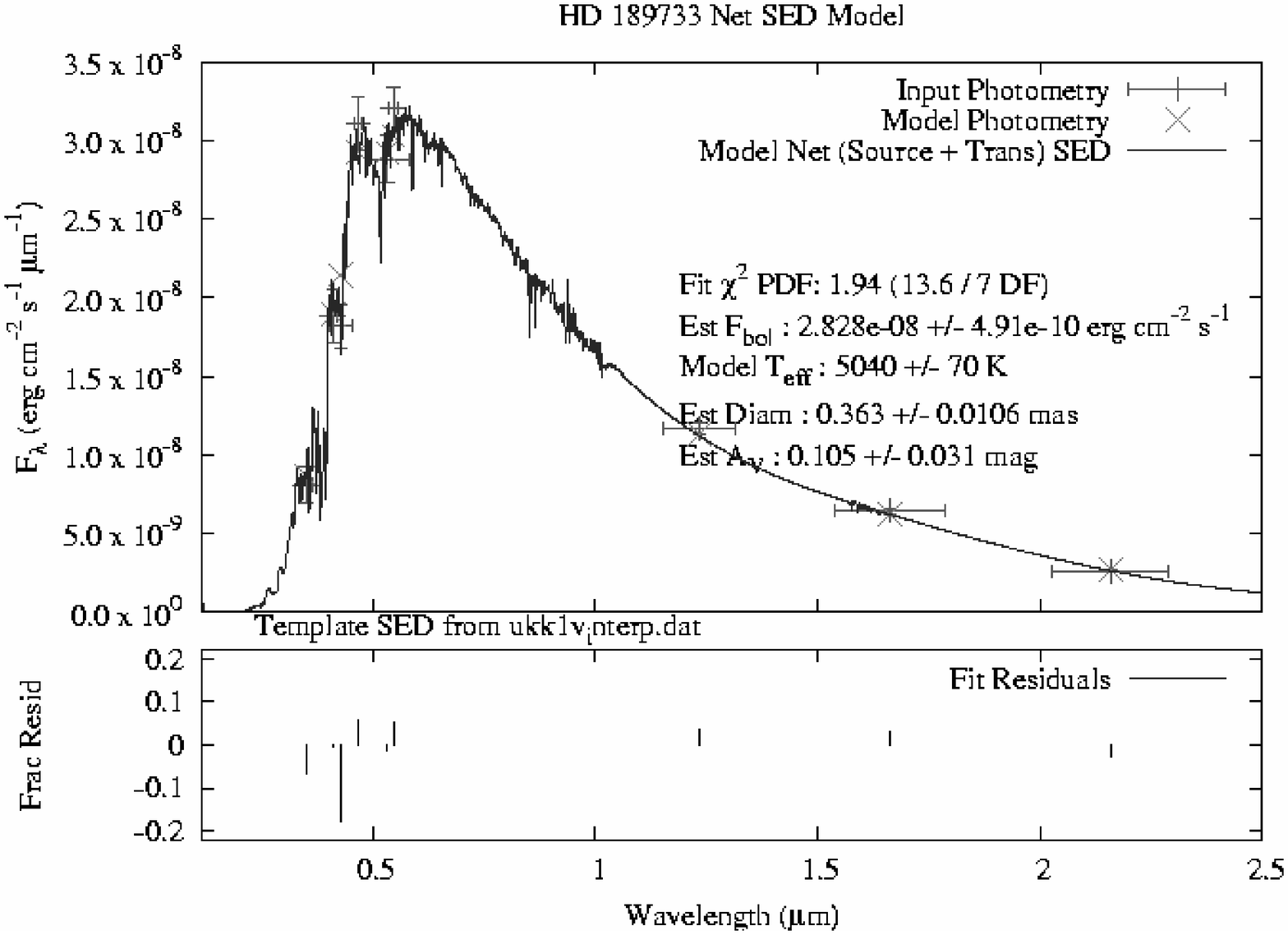}\\
  \includegraphics[width=0.5\textwidth]
  {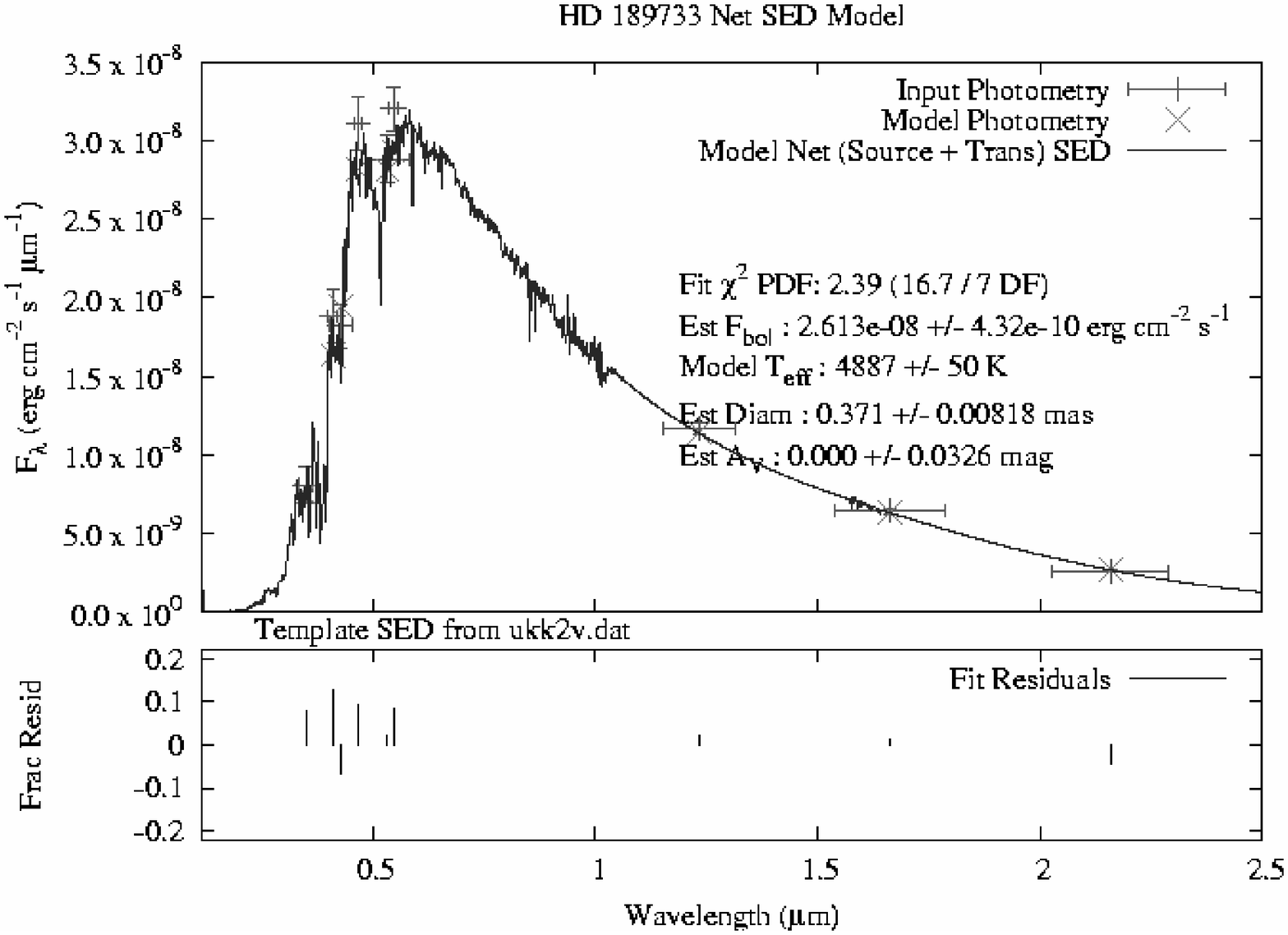}
  \caption{SED fits of HD 189733. The top panel shows the photometric fit for a K0~V, the middle panel shows the K1~V fit, and the bottom panel shows the K2~V fit. All three fits allowed the reddening factor to vary.}
  \label{HD189733_SEDs}
\end{figure}

To explore what is a more appropriate error for an inferred angular size, we executed a spectral energy distribution fit of the available spectrophotometry for HD~189733 cited above.  Given the known variability of HD~189733, the quoted millimagnitude error estimates of the Tycho and Str{\"o}mgren photometric data points were increased by a factor of ten. These photometric data points were fit to the solar-abundance K0~V and K2~V templates available from \citet{1998PASP..110..863P}, with the resulting fit values for reddening, bolometric flux, and angular diameter seen in Table \ref{table_SEDfit}, along with the appropriate $\chi^2$ per degree of freedom ($\chi^2$ PDF) values. Unfortunately, a K1~V template is not available in that library, although an estimate of one from interpolating between the two bracketing spectral types was synthesized by us for testing this spectral type. These fits are seen in Figure \ref{HD189733_SEDs}.

The appropriate model spectrum from \citet{2005A&A...442.1127M} for a 5000~K star was fit with a $\chi^2$ PDF of 2.80, but this model (nor any other available in the literature) unfortunately only covered the visible portion of the spectrum. The 115 to 2500 nm range of \citet{1998PASP..110..863P} was necessary to fully characterize the available photometry, and thus we constrained our analysis to this particular set of stellar templates.

Our finding is that, even with this highly detailed analysis of the stellar spectral energy distribution, the most appropriate modelling of that SED reveals a predicted angular size of only $\theta = 0.363 \pm 0.011$ mas - a 3\% error bar - which corresponds to a stellar linear radius of $R = 0.752 \pm 0.026$ $R_\odot$.

\section{Discussion}
Our new direct determination for the angular diameter of HD~189733 of $\Theta_{\rm LD} = 0.377 \pm 0.024$~mas can be combined with the Hipparcos parallax for the star of $\pi = 51.9 \pm 0.9$~mas \citep{1997A&A...323L..49P} to give a physical radius for the star of $R_{\rm star} = 0.779 \pm 0.052$~$R_\odot$, which is about 3\% larger than that adopted by \citet{2006ApJ...650.1160B}.  

By the nature of the light curve analysis, the relative increase for the radius of the host star will directly translate into the same relative increase in the radius of the planet HD~189733b. Thus, revising the radius of \citet{2006ApJ...650.1160B} of 1.154~$R_{\rm Jup}$, our new estimate for this value is $R_{\rm planet} = 1.19 \pm 0.08$~$R_{\rm Jup}$. Furthermore, adopting the value of \citet{2005A&A...444L..15B} for the mass of the planet of 1.15~$M_{\rm Jup}$, we derive a new estimate for the density of HD~189733b to be $\rho = 0.91 \pm 0.18$~g~cm$^{-3}$. These values are in good agreement with \citet{winn2007}, who used transit photometry to constrain the stellar and planetary radii. The values of $M_{\rm planet}$, $R_{\rm planet}$, and $\rho_{\rm planet}$ are all consistent with the modest collection of these parameters presently available for transiting exoplanet systems and support the conclusion that HD~189733 is not among the few hot Jupiters that present extraordinarily large radii for their masses.

\section{Interferometric Non-Detection of Binarity of HD 189733}

Given the higher resolution of interferometric arrays, a possible close-separation tertiary companion may affect our measures of HD~189733's visibility and thereby complicate our interpretation. As such, it was prudent for us to also observe HD~189733 with the Palomar Testbed Interferometer (PTI) \citep{1999ApJ...510..505C}, an instrument with intermediate baselines on a variety of sky projections, suitable for exploration of possible unseen nearby luminous (stellar) companions. PTI has been demonstrated to be sensitive to nearby companions with $\Delta K < 4.0$ \citep{1998ApJ...504L..39B}, which for a K2-3~V primary star rules out any M-dwarf companions \citep{1988PASP..100.1134B}.

PTI observed HD~189733 in the $K$-band on the nights of 2006 June 10-12, 2006 June 24, and 2006 July 8-10. Four of those nights used PTI's 85-m NW baseline configuration, two used the 110-m NS
baseline, and one night was a 85-m SW baseline night. For all of these nights, HD~189733's normalized $V$ data points were indistinguishable from unit visibility, which corresponds to a completely unresolved point source, as would be expected for a single $\sim 0.37$ mas star being observed by PTI at 2.2~$\mu$m.

\section{Conclusion}

Our results for the radii of the host star and planet in the HD 189733 exoplanet system are formally in good agreement with existing measurements of these parameters as well as with the estimate for the density of the planet and have the additional and significant merit that they represent $direct$ measurements of stellar and planetary diameters that do not rely upon inferences about stellar atmospheres. While the diameter measurements are currently at a 6\% level of accuracy, we expect to improve this considerably as we implement fringe detection at shorter wavelengths at the CHARA Array. In the meantime, these results demonstrate a new role that long-baseline optical/infrared interferometry can play in the field of exoplanet astronomy.

\acknowledgements

We would like to thank Andy Boden for sharing his SEDfit tools with us, which we used to produce the fits seen in Figures 1 and 3, and we appreciate the care that CHARA Array Operator P.J. Goldfinger used in obtaining many of these observations. The CHARA Array is funded by the National Science Foundation through NSF grants AST-0307562 and AST-06006958 and by Georgia State University through the College of Arts and Sciences and the Office of the Vice President for Research. Observations with PTI are made possible through the efforts of the PTI Collaboration, which we acknowledge. Funding for PTI was provided to the Jet Propulsion Laboratory under its TOPS (Towards Other Planetary Systems), ASEPS (Astronomical Studies of Extrasolar Planetary Systems), and Origins programs and from the JPL Director's Discretionary Fund. Part of the work described in this paper was performed at the Jet Propulsion Laboratory under contract with the National Aeronautics and Space Administration. This research has made use of the SIMBAD literature database, operated at CDS, Strasbourg, France, and of NASA's Astrophysics Data System. This publication makes use of data products from the Two Micron All Sky Survey, which is a joint project of the University of Massachusetts and the Infrared Processing and Analysis Center/California Institute of Technology, funded by the National Aeronautics and Space Administration and the National Science Foundation.

\clearpage

\begin{deluxetable}{cccc}
\tablewidth{0pc} \tablecaption{Interferometric Measurements of HD~189733.\label{visibility}}
\tablehead{
 \colhead{MJD}        & \colhead{Baseline} & \colhead{Visibility} & \colhead{$\sigma$Vis} \\
 \colhead{(53886.0 +)} & \colhead{(m)}      & \colhead{ }          & \colhead{ } \\
}
\startdata
0.905     &         330.5         &        0.851    &    0.071\\
1.936     &         327.9         &        0.843    &    0.056\\
1.958     &         324.9         &        0.857    &    0.054\\
8.865     &         330.5         &        0.869    &    0.034\\
76.742    &         326.5         &        0.909    &    0.069\\
76.761    &         323.8         &        0.863    &    0.049\\
76.778    &         321.3         &        0.877    &    0.045\\
76.793    &         319.0         &        0.839    &    0.045\\
76.824    &         315.5         &        0.829    &    0.061\\	
\enddata
\end{deluxetable}

\clearpage

\begin{deluxetable}{cc|cccc}
\tablecolumns{6}
\tabletypesize{\scriptsize}
\tablewidth{0pc}
\tablecaption{Spectral energy distribution fits for HD~189733 photometry to 
empirical spectral templates.\label{table_SEDfit}}
\tablehead{
\multicolumn{ 2}{c}{Model Parameters\tablenotemark{a}} & 
\multicolumn{ 4}{c}{Fitted parameters}\\
\colhead{Spectral type} & \colhead{Effective Temperature} & \colhead{$\chi^2$ PDF}
   &  \colhead{Reddening} & \colhead{Bolometric Flux $F_{\rm BOL}$} & 
\colhead{Angular Diameter} \\
   & \colhead{$T_{\rm EFF}$ (K)} &
   &  \colhead{$A_V$ (mag)} & \colhead{(erg cm$^{-2}$ s$^{-1}$ $\mu$m$^{-1}$)}
   & \colhead{$\theta$ (mas)} \\
}
\startdata
K0~V & $5188 \pm 50$ & 2.62 & $0.308 \pm 0.031$ & $3.207 \pm 0.061\times 10^{-8} $ & $0.365 \pm 0.008$ \\
K1~V\tablenotemark{b} & $5040 \pm 70$ & 1.94 & $0.105 \pm 0.031$ & $2.828 \pm 0.049 \times 10^{-8}$ & $0.363 \pm 0.011$ \\
K2~V & $4887 \pm 50$ & 2.39 & $0.00 \pm 0.030$ & $2.613 \pm 0.043 \times 10^{-8}$ & $0.371 \pm 0.008$ \\
\enddata
\tablenotetext{a}{Models from \citet{1998PASP..110..863P}.}
\tablenotetext{b}{K1~V model interpolated from the K0~V and K2~V models.}
\end{deluxetable}

\clearpage

\clearpage

\clearpage

\begin{thebibliography}{99}

\bibitem[Bakos et al.(2006a)]{2006ApJ...641L..57B} Bakos, G.~{\'A}., et al.\ 2006a, \apjl, 641, L57
\bibitem[Bakos et al.(2006b)]{2006ApJ...650.1160B} Bakos, G.~{\'A}., et al.\ 2006b, \apj, 650, 1160
\bibitem[Baraffe et al.(1998)]{1998A&A...337..403B} Baraffe, I., et al.\ 1998, \aap, 337, 403
\bibitem[Bessell(2000)]{2000PASP..112..961B} Bessell, M.~S.\ 2000, \pasp, 112, 961
\bibitem[Bessell \& Brett(1988)]{1988PASP..100.1134B} Bessell, M.~S., \& Brett, J.~M.\ 1988, \pasp, 100, 1134
\bibitem[Boden et al.(1998)]{1998ApJ...504L..39B} Boden, A.~F., et al.\ 1998, \apjl, 504, L39
\bibitem[Bouchy et al.(2005)]{2005A&A...444L..15B} Bouchy, F., et al.\ 2005, \aap, 444, L15
\bibitem[Charbonneau et al.(2000)]{2000ApJ...529L..45C} Charbonneau, D., et al.\ 2000, \apjl, 529, L45
\bibitem[Claret et al.(1995)]{1995A&AS..114..247C} Claret, A., Diaz-Cordoves, J., \& Gimenez, A.\ 1995, \aaps, 114, 247
\bibitem[Colavita et al.(1999)]{1999ApJ...510..505C} Colavita, M.~M., et al.\ 1999, \apj, 510, 505
\bibitem[Cutri et al.(2003)]{2003tmc..book.....C} Cutri, R.~M., et al.\ 2003, 
The IRSA 2MASS All-Sky Point Source Catalog, NASA/IPAC Infrared Science 
Archive.~http://irsa.ipac.caltech.edu/applications/Gator/
\bibitem[Girardi et al.(2002)]{2002A&A...391..195G} Girardi, L., et al.\ 2002, \aap, 391, 195
\bibitem[Hanbury Brown et al.(1974)]{1974MNRAS.167..475B} Hanbury Brown, R., et al.\ 1974, \mnras, 167, 475
\bibitem[Kervella et al.(2004)]{2004A&A...426..297K} Kervella, P., et al.\ 2004, \aap, 426, 297
\bibitem[Masana et al.(2006)]{2006A&A...450..735M} Masana, E., Jordi, C., \& Ribas, I.\ 2006, \aap, 450, 735
\bibitem[McAlister et al.(2005)]{2005ApJ...628..439M} McAlister, H.~A., et al.\ 2005, \apj, 628, 439
\bibitem[Munari et al.(2005)]{2005A&A...442.1127M} Munari, U., et al.\ 2005, \aap, 442, 1127
\bibitem[Olsen(1993)]{1993A&AS..102...89O} Olsen, E.~H.\ 1993, \aaps, 102, 89
\bibitem[Perryman et al.(1997)]{1997A&A...323L..49P} Perryman, M.~A.~C., et al.\ 1997, \aap, 323, L49
\bibitem[Pickles(1998)]{1998PASP..110..863P} Pickles, A.~J.\ 1998, \pasp, 110, 863
\bibitem[Rowe(2007)]{rowe2007} Rowe, J.\ 2007, private communication
\bibitem[Sato et al.(2005)]{2005ApJ...633..465S} Sato, B., et al.\ 2005, \apj, 633, 465
\bibitem[ten Brummelaar et al.(2005)]{2005ApJ...628..453T} ten Brummelaar, T.~A., et al.\ 2005, \apj, 628, 453
\bibitem[van Belle \& van Belle(2005)]{2005PASP..117.1263V} van Belle, G.~T. \& van Belle, G.\ 2005, \pasp, 117, 1263
\bibitem[Winn et al.(2007)]{winn2007} Winn, J.~N., et al.\ 2007, \aj, in press
\end{thebibliography}
\end{document}